\documentclass[aps,prd,reprint,preprintnumbers,superscriptaddress,showpacs,twocolumn]{revtex4-1}
\usepackage{latexsym,graphicx,amssymb,amsmath,mathrsfs}
\usepackage{setspace,bm}
\usepackage[breaklinks, colorlinks=true, pdfstartview=FitV, linkcolor=red, citecolor=blue, urlcolor=blue]{hyperref}
\usepackage[usenames]{color}
\usepackage{latexsym}
\usepackage{epstopdf}
\usepackage{mathtools}
\usepackage{url}
\usepackage{comment}
\usepackage{braket}
\usepackage{CJKutf8}

\begin{document}

\title{Polyakov-loop phase, Roberge-Weiss periodicity and thermodynamics}

\author{Kouji Kashiwa}
\email[]{kashiwa@fit.ac.jp}
\affiliation{Fukuoka Institute of Technology, Wajiro, Fukuoka 811-0295, Japan}

\author{Hiroaki Kouno}
\affiliation{Department of Physics, Saga University, Saga 840-8502, Japan}

\begin{abstract}
In this paper, we discuss the role of Roberge-Weiss periodicity in the thermodynamics of quantum chromodynamics at moderately high temperature, where the semi-quark-gluon plasma is expected.
From the construction of the grand canonical partition function at zero and also at finite density via the canonical approach, we are able to discuss the relation between contributions of the Polyakov-loop phase and Roberge-Weiss periodicity.
Then, we can conclude that the existence of Roberge-Weiss periodicity is a necessary condition to reproduce exact results at moderately high temperature.
\end{abstract}
\maketitle

\section{Introduction}

Exploring the non-perturbative properties of quantum chromodynamics (QCD) is an interesting and important subject to understand our universe; see Ref.\,\cite{Fukushima:2010bq} as an example.
At finite temperature ($T$) but zero real chemical potential ($\mu_\mathrm{R}=0$), we can perform the lattice QCD simulation, and then non-perturbative properties can be exactly discussed by using the lattice QCD simulation in principle.
However, at finite $\mu_\mathrm{R}$, QCD has the sign problem, and thus some methods are necessary to control the sign problem; see Ref.\,\cite{deForcrand:2010ys,Nagata:2021ugx} as an example.

Non-physical systems, such as the imaginary chemical potential ($\mu_\mathrm{I}$) region, have a strong impact on realistic systems, that is, the real chemical potential region; for example, see Ref.\,\cite{deForcrand:2010ys,Kashiwa:2019ihm}.
Actually, we can construct the grand canonical partition function at finite $\mu_\mathrm{R}$ starting from finite $\mu_\mathrm{I}$~\cite{Roberge:1986mm}.
In this paper, we consider the problem that Roberge-Weiss (RW) periodicity is absent in the calculation at moderately high $T$; the Feynman diagrammatic approach, such as the hard thermal loop perturbation theory~\cite{Haque:2014rua}, the thermal Dyson-Schwinger equation approach~\cite{Roberts:2000aa,Maas:2005hs} and the functional renormalization group approach for the effective potential~\cite{Wetterich:1992yh}, sometimes has the same problem.
The term "moderately high" means that the Polyakov loop has not reached one, but it is somewhat significant.
We discuss the consequence of the absence of RW periodicity~\cite{Roberge:1986mm} at finite $\mu_\mathrm{I}$ for the thermodynamics at finite $\mu_\mathrm{R}$.

In the canonical approach~\cite{Alexandru:2005ix,deForcrand:2006ec,Fukuda:2015mva,Bornyakov:2016wld,Oka:2017kny,Wakayama:2018wkc}, we start with the imaginary chemical potential~\cite{Roberge:1986mm,Hasenfratz:1991ax}.
This means that we can construct the canonical partition function with a fixed quark number using the grand canonical partition function at finite $\mu_\mathrm{I}$.
Since the process uses the Fourier transformation, this process can also be reversed.
This indicates that we can examine the validity of the analytical method via the imaginary chemical potential region. 
Actually, there are several interesting and important properties of QCD at finite $\mu_\mathrm{I}$ such as RW periodicity and RW transition~\cite{Roberge:1986mm}.
Furthermore, with careful investigation of the canonical partition function using the canonical approach, several phases are clarified from the structure of the canonical sector~\cite{Kashiwa:2019dqn} and different types of probability distribution~\cite{Kashiwa:2021toz}.
Since the canonical partition function can provide important information about QCD, it is interesting to discuss QCD properties from the point of view of the canonical sector.

This paper is organized as follows.
In the next section, we briefly explain the properties of QCD at finite $\mu_\mathrm{I}$.
Section\,\ref{Sec:Canonical_method} explain the canonical approach.
Section~\ref{sec:Discussions} shows some discussions of the Feynman diagrammatic approach and the consequence of the absence of the Roberge-Weiss transition.
Section\,\ref{Eq:Summary} is devoted to the summary of this research.

\section{Imaginary chemical potential}

At finite $\mu_\mathrm{I}$, it is known that QCD has the following interesting properties.
Since the properties play a crucial role in this study, we will briefly summarize them in the following:
\begin{description}
    \item[RW periodicity]
    Several thermodynamic quantities and order parameters have the  $2\pi/N_\mathrm{c}$ periodicity along the $\theta$-axis where $\theta$ is the dimensionless imaginary chemical potential, $\theta \equiv \mu_\mathrm{I}/T$, and $N_\mathrm{c}$ is the number of colors.
    This periodicity is called RW periodicity, and it can be proven model-independently \cite{Roberge:1986mm}.
    \item[RW transition]
    RW periodicity is induced by different mechanisms at low and high temperatures; i.e. the balance between gluon and quark contributions.
    At $\theta=(2k-1)\pi/N_\mathrm{c}$ with $k \in \mathbb{Z}$, quantities have singularities at high $T$.
    This singularity originates from $\mathbb{Z}_{N_\mathrm{c}}$ replicas; see Fig.\,2 in Ref.\,\cite{Roberge:1986mm} and Fig.\,7 in Ref.\,\cite{Kouno:2009bm} as an example.
    This means that a particular replica has the smallest thermodynamic potential at a certain $\theta$ and then becomes a physical state.
    The lowest state changes with changing $\theta$ and two $\mathbb{Z}_{N_\mathrm{c}}$ replicas are degenerated at $\theta = (2k-1) \pi /N_\mathrm{c}$; it is the origin of the RW transition.    
    The $\theta$-odd and -even quantities have first-order (gap) and second-order (cusp) singularities, and they can co-exist; see Ref.\,\cite{Kashiwa:2009zz} for details of the mechanism of co-existing. 
    These characterize the phase transition, which is called the RW transition.
    \item[RW endpoint]
    Quantities oscillate gently and smoothly along the $\theta$-axis at low $T$, but have singularities along the $\theta$-axis at high $T$.
    Due to the difference, there should be an endpoint of the first-order RW transition line at the critical value of $T$.
    This endpoint is the so-called RW endpoint.
    This point may characterize the crossover behavior at $\mu=0$.
\end{description}
For details of the above properties, see Refs.\,\cite{Roberge:1986mm,Kashiwa:2019ihm}.

At moderately high $T$, the Polyakov loop does not reach one, but RW transition exists along the $\theta$-axis.
In this region, the system may be the semi-quark-gluon plasma; for details of the semi-quark-gluon plasma, see Refs.\,\cite{Hidaka:2009hs,Hidaka:2009ma} as an example. 
This region is the main focus of this paper.

\section{Canonical approach}
\label{Sec:Canonical_method}

It is well known that the canonical partition function (${\cal Z}_\mathrm{C}$) with fixed quark number ($k$) can be constructed by using the grand canonical partition function (${\cal Z}_\mathrm{GC}$) at finite $\theta$ as
\begin{align}
    {\cal Z}_\mathrm{C} (Q)
    &= \sum_{n=-\infty}^\infty \bra{n} e^{-\beta \hat{\cal H}} \delta (\hat{N}-Q) \ket{n} 
\nonumber\\
    &= \frac{1}{2\pi} \int_{-\pi}^\pi
       e^{i Q \theta} {\cal Z}_\mathrm{GC} (\theta) \, d\theta,
\label{Eq:canonical}
\end{align}
where $\hat{\cal H}$ means the Hamiltonian, $n \in \mathbb{Z}$ mean eigenvalues of the quark number operator $\hat{N}$ and
$Q \in \mathbb{Z}$.
In this paper, we consider the sufficiently large but finite spatial volume.
This formulation is already applied to the lattice QCD~\cite{Alexandru:2005ix,deForcrand:2006ec,Fukuda:2015mva,Bornyakov:2016wld,Oka:2017kny,Wakayama:2018wkc} and QCD effective models~\cite{Morita:2015tma,Wakayama:2020dzz,Kashiwa:2021czl,Kashiwa:2021til,Kashiwa:2021toz}.
The canonical partition function, of course, obeys a canonical ensemble.
Using the partition function, we can evaluate several quantities.

With the fugacity expansion, we obtain the grand canonical partition function at finite $\mu_\mathrm{R}$ as
\begin{align}
    {\cal Z}_\mathrm{GC}(\mu_\mathrm{R})
    &= \sum_{Q=-\infty}^\infty
       \exp \Bigl( {Q \frac{\mu_\mathrm{R}}{T}} \Bigr)
       {\cal Z}_\mathrm{C}(Q)
       \nonumber\\
    &= \sum_{Q=-\infty}^\infty
       \xi^Q
       {\cal Z}_\mathrm{C}(Q),
\label{Eq:fugacity}
\end{align}
where $\xi$ is the so called fugacity.
It is noted that the RW periodicity is not explicitly included here because we are also interested in the case without it.
The inclusion of the periodicity is discussed in Sec.\,\ref{sec:RWP}.
In the canonical approach, we usually discuss the grand canonical partition function and related quantities, but the canonical partition function itself is interesting: The canonical partition functions relate to the multiplicity distribution that can be extracted from collision experiments; see Refs.\,\cite{STAR:2010mib,Luo:2012kja} as the experimental data and the recent review~\cite{Fukushima:2020yzx}.
Actually, such a fact was used to combine lattice QCD data and experimental data at finite density through Lee-Yang zeros~\cite{yang1952statistical,lee1952statistical}; see Ref.\,\cite{Nakamura:2013ska} as an example.

\subsection{Roberge-Weiss periodic case}
\label{sec:RWP}

When RW periodicity exists, the canonical partition function (\ref{Eq:canonical}) with fixed quark number $Q$ can be constructed by using ${\cal Z}_\mathrm{GC}$ at finite $T$ and $\theta$ as
\begin{align}
    &{\cal Z}_\mathrm{C} (Q)
    \nonumber\\
    &= \frac{1 + z^{Q} + z^{2Q}}{2\pi}
       \int_{-\pi/3}^{\pi/3}
       e^{i Q \theta} {\cal Z}_\mathrm{GC} (\theta) \, d\theta
       \nonumber\\
    &=
    \begin{dcases}
    \frac{3}{2\pi}  
    \int_{-\pi/3}^{\pi/3}
       e^{i Q \theta} {\cal Z}_\mathrm{GC} (\theta) \, d\theta
       & (Q= 3 k)\\
       ~0 & (Q \neq 3 k)
    \end{dcases}
    ,
\label{Eq:canonical_RW_p}
\end{align}
where $k \in \mathbb{Z}$ and
\begin{align}
    z = \exp \Bigl( \frac{2\pi i}{3} \Bigr),
\end{align}
is the ${\mathbb Z}_3$ factor.
From here, we set $N_\mathrm{c} = 3$ and do not explicitly show $T$ for the argument of the partition function because we are interested in the $\mu$-dependence with fixed $T$.

With the fugacity expansion, we have the grand canonical partition function as
\begin{align}
    {\cal Z}^{\mathrm{RW}}_\mathrm{GC}(\mu_\mathrm{R})
    &= \sum_{k=-\infty}^\infty
       \exp \Bigl( 3 k \frac{\mu_\mathrm{R}}{T} \Bigr)
       {\cal Z}_\mathrm{C}(3 k)
       \nonumber\\
    &= \sum_{k=-\infty}^\infty
       \exp \Bigl( Q_\mathrm{B} \frac{\mu_\mathrm{R}}{T} \Bigr)
       {\cal Z}_\mathrm{C}(Q_\mathrm{B}),
\label{Eq:fugacity_RW}
\end{align}
where $Q_\mathrm{B} = 3k$ and we use the well known fact that $N_\mathrm{c}$ multiples of $n$ only contribute ${\cal Z}_\mathrm{C}$ because of RW periodicity as shown in Eq.\,(\ref{Eq:canonical_RW_p}); see also Ref.\,\cite{Roberge:1986mm}.
In the above, we start from ${\cal Z}_\mathrm{GC}(\theta)$ to obtain ${\cal Z}_\mathrm{C}(Q)$, but we can reverse it.
This means that the canonical partition function is not correct if RW transition cannot be reproduced in the calculation; this is the most important fact in this paper.
We discuss more details in the following.

\subsection{Roberge-Weiss non-periodic case}

In the RW non-periodic approach, which does not have RW transition, the canonical partition function (\ref{Eq:canonical}) becomes
\begin{align}
    {\cal Z}_\mathrm{C} (Q)
    &= \frac{1}{2\pi}
       \int_{-\pi/3}^{\pi/3}
       e^{i Q \theta}
       \tilde{\cal Z}_\mathrm{GC}
       \,d\theta,
\label{Eq:canonical_RW}
\end{align}
where
\begin{align}
&\tilde{\cal Z}_\mathrm{GC} (\theta )
\nonumber\\
&= {\cal Z}_\mathrm{GC} ( \theta )
 + z^{Q} {\cal Z}_\mathrm{GC} \Bigl( \theta + \frac{2\pi}{3}\Bigr)
 + z^{2Q} {\cal Z}_\mathrm{GC} \Bigl( \theta + \frac{4 \pi}{3}\Bigr)
\nonumber\\
&\neq \Bigl[ 1 + z^{Q} + z^{2Q} \Bigr]
   {\cal Z}_\mathrm{GC}( \theta ),
\label{eq:inequality}
\end{align}
here the third line is corresponding to the RW periodic case.
It should be noted that $\tilde{\cal Z}_\mathrm{GC}(\theta)$ must have, at least, the trivial $2 \pi$ periodicity due to the QCD foundation; the imaginary dimensionless chemical potential $\theta$ is known to be able to be treated as the temporal boundary condition of quarks~\cite{Bilgici:2008qy}.
The inequality (\ref{eq:inequality}) is the main point of the paper, since it indicates the reason why the Feynman diagrammatic approach, which does not have RW periodicity, may deviate from the exact result at moderately high $T$ from the point of view of the canonical approach, as shown later.

With the fugacity expansion, we have the grand canonical partition function as
\begin{align}
    {\cal Z}^{\mathrm{nonRW}}_\mathrm{GC}(\mu_\mathrm{R})
    &= \sum_{Q=-\infty}^\infty
       \exp \Bigl( {Q \frac{\mu_\mathrm{R}}{T}} \Bigr)
       {\cal Z}_\mathrm{C}(Q).
\label{Eq:fugacity_np}
\end{align}
Since there is no RW transition, $(Q\, \mathrm{mod}\,3) \neq 0$ contributions remain unlike the RW periodic case.

\section{Discussions}
\label{sec:Discussions}

In this section, we discuss the consequence of the absence of the RW periodicity and the transition at moderately high $T$.
We used the Fourier decomposition to make simple discussions.

\subsection{Fourier decomposition}

At high $T$, physical quantities, such as partition function, entropy, density, etc., should oscillate gently and smoothly at finite $\theta$ on a particular replica; RW transition, which is a singularity, is induced by the transfer from the replica to another replica.
Thus, we can assume the form of ${\cal Z}_\mathrm{GC}(\theta)$ as
\begin{align}
    {\cal Z}_\mathrm{GC}(\theta) &= \sum_{m=-m_\mathrm{max}}^{m_\mathrm{max}} a_m \cos(m \theta),
\end{align}
where $m \in \mathbb{Z}$, $m_\mathrm{max}$ means the cutoff of the sum which is basically $+\infty$ and $a_m \in \mathbb{R}$ mean Fourier coefficients.
If the form of the oscillation behavior is gentle and smooth, $m_\mathrm{mas}$ can be set as a small number.
This functional form only has the trivial $2\pi$ periodicity, but RW periodicity can be reproduced by considering its $\mathbb{Z}_{N_\mathrm{c}}$ replicas.

After performing the integration in Eq.\,(\ref{Eq:canonical}), $m=Q$ terms are only survived in the canonical partition function due to the properties of Fourier decomposition.
Therefore, by taking into account the properties of the RW periodicity\,(\ref{Eq:canonical_RW_p}), we obtain the grand canonical partition functions as
\begin{align}
    {\cal Z}^{\mathrm{RW}}_\mathrm{GC}(\mu_\mathrm{R})
    &= \frac{1}{2} \sum_{k=-m_\mathrm{max}}^{m_\mathrm{max}}
       \exp \Bigl( {3k \frac{\mu_\mathrm{R}}{T}} \Bigr)
       a_{3 k}
    \nonumber\\
    &= \frac{1}{2} \sum_{k=-m_\mathrm{max}}^{m_\mathrm{max}}
       \exp \Bigl( {Q_\mathrm{B} \frac{\mu_\mathrm{R}}{T}} \Bigr)
       a_{Q_\mathrm{B}}
    ,
\label{Eq:Zc_RW}
\end{align}
for the RW periodic approaches and
\begin{align}
    {\cal Z}^{\mathrm{nonRW}}_\mathrm{GC}(\mu_\mathrm{R})
    &= \frac{1}{2} \sum_{Q=-3m_\mathrm{max}}^{3m_\mathrm{max}}
       \exp \Bigl( {Q \frac{\mu_\mathrm{R}}{T}} \Bigr)
       a_{Q},
\label{Eq:Zc_trivial}
\end{align}
for the RW non-periodic approaches.
In the RW non-periodic approaches, contributions from all possible values of $Q$ survive, unlike in the RW periodic approaches.
Then, we have
\begin{align}
    {\cal Z}^{\mathrm{RW}}_\mathrm{GC}(\mu_\mathrm{R})
     \neq
    {\cal Z}^{\mathrm {nonRW}}_\mathrm{GC}(\mu_\mathrm{R}),
    \label{eq:enq}
\end{align}
because $Q_\mathrm{B}=3k$.
The $(Q\,\mathrm{mod}\,3)$ contributions match each other, but there are additional contributions in Eq.\,(\ref{Eq:Zc_trivial}).
This difference between Eq.\,(\ref{Eq:Zc_RW}) and (\ref{Eq:Zc_trivial}) is the consequence of the contributions of the Polyakov-loop phase contributions.
Since the grand canonical partition function generates thermodynamic quantities and some other observables, the mismatch between the RW periodic and nonperiodic cases is also true for thermodynamic quantities and some other observables.
This point may be important when we consider the external magnetic field to evaluate the Polyakov loop in the non-lattice calculation; see Appendix~\ref{sec:A_PL}.

\subsection{Feynman diagrammatic approach}
In the Feynman diagrammatic approach, we first prepare propagators and vertices, and then we make Feynman diagrams.
For example, we calculate loop diagrams with coupling counting in the perturbation theory, the one-loop diagram in the functional renomarlization group approach, and so on.
However, we need the gauge fixing to make propagators and vertices in the Feynman diagrammatic approach.
Then, RW periodicity disappears if we do not take special care because we set the expansion point of the perturbation in the trivial $\mathbb{Z}_{N_\mathrm{c}}$ sector.
This indicated that we cannot have a nontrivial sector of holonomy for QCD because the Polyakov loop ($\Phi$) always takes a real value;
\begin{align}
    \Phi(\vec{x}) &= \frac{1}{3} \mathrm{tr} \exp \Bigl[ ig \int d \tau \, {\cal A}_\mathrm{tri}(x) \Bigr] \in \mathbb{R},
\end{align}
where ${\cal A}_\mathrm{tri}$ is the gauge field fixed in the trivial $\mathbb{Z}_{N_\mathrm{c}}$ sector.
This means that we cannot reproduce RW periodicity if we generate configurations after gauge fixing because the configurations correspond to ${\cal A}_\mathrm{tri}$.
In the lattice QCD simulation, the gauge fixing is usually imposed after the configuration generation, if it is needed, and thus RW periodicity can exist because we do not fix the expansion point in this way.
These facts may induce a deviation of the RW-nonperiodic Feynman diagrammatic approach from the exact results, as explained in the following.

If we remain related quantities with the phase of the Polyakov loop in the effective potential in the construction part of the perturbation theory, we can reproduce RW transition: see Ref.\,\cite{Roberge:1986mm}.
Unfortunately, it is not the standard way in the Feynman diagrammatic approach.
A similar situation occurs in the functional renormalization group method: we should carefully introduce the quantities to the Matsubara frequency~\cite{Braun:2009gm}.
These results indicate that we may miss some contributions of Polyakov-loop phase in the simple Feynman diagrammatic approach.
However, in previous studies, there were no detailed discussions of what happens in thermodynamics when RW periodicity is not manifested.
Actually, this point is now clarified by Eqs.\,(\ref{eq:inequality}), (\ref{Eq:Zc_RW}) $\sim$ (\ref{eq:enq}).

To include RW periodicity in the Feynman diagrammatic approach, one possible way is using the semi-classical expansion of the gauge field~\cite{Hidaka:2009ma,Hidaka:2009hs} that the gluon field (${\cal A}$) is decomposed as the sum of the classical field ($A$) and the fluctuation (${\cal B}$).
Then, the loop diagrams are estimated with a suitable setup of the classical field.
The classical field $A$ manifests $\mathrm{tr} A = 0$ in the color space, and the propagator are obtained from the bilinear of ${\cal B}$.
This point can be understood from the background gauge field method when holonomy plays a important role;
\begin{align}
    \int {\cal D} A \int {\cal D} {\cal B} \, e^{-S_\mathrm{E}[A,B^\Lambda]}
    \neq
    \int {\cal D} {\cal A} \, e^{-S_\mathrm{E}[{\cal A}^\Lambda]},
    \label{eq:background_field_method}
\end{align}
where ${\cal A}^\Lambda$ and ${\cal B}^\Lambda$ are gauge fixed fields, and $A$ is determined from the equation of motion or minimizing the effective potential because it is a classical field.
Then, the integration by $A$ may be replaced by the sum of solutions and becomes finite.
It should be noted that inequality in Eq.\,(\ref{eq:background_field_method}) can become equality when the trivial center only contributes to the integral.
A similar problem also appears if we consider the chemical potential in the lattice quantum electrodynamics (QED) on the torus.
Then, we can introduce the toron field that appears from the decomposition of the photon field to recover the correct behavior of QED; for example, see Ref.\,\cite{Narayanan:2012du,Tanizaki:2016xcu} and references therein.
The toron field is directly related to the holonomy of QED and thus works somewhat similarly to $A$ in Eq.\,(\ref{eq:background_field_method}).
It is noted that there are discussions suggesting that the RW transition is caused by the massless nature of quarks \cite{Rogalyov:2024ezm}.
However, this topic is out of the scope of the current study, which considers realistic QCD at moderate $T$.
Therefore, we continue to keep it as our future work.

The simplest setup to recover RW periodicity is to just consider the classical field in the Matsubara frequency with the semi-classical expansion as
\begin{align}
    \omega_n = (2n+1) \pi T + i \mu_\mathrm{R} + A_4,
\label{eq:extension}
\end{align}
where $\omega_n$ is the Matsubara frequency with $n \in \mathbb{Z}$.
Then, $A_4$ can lead the nontrivial holonomy to the calculation because it is not yet fixed.
To make it work well even at finite $\mu_\mathrm{I}$, we need an additional potential term originating from the classical field for the semi-quark-gluon plasma, but this is not yet clear; this additional potential term controls the $T$-dependence of the Polyakov loop.
Therefore, the classical field can cure the problem induced by gauge fixing.
It is noted that the simplest treatment above (\ref{eq:extension}) without an additional potential term can recover RW periodicity because $i \mu + A_4$ is the RW periodic combination; it relates to the extended $\mathbb{Z}_{N_\mathrm{c}}$ symmetry~\cite{Sakai:2008ga}.
RW transition can also be reproduced with Eq.\,(\ref{eq:extension}), but there are differences with the exact result, particularly near $\theta = \pi/N_\mathrm{c}$; see Appendix~\ref{sec:examples}.
Thus, if we do not introduce a suitable distribution of the classical field, the result should deviate from the exact result.
Such a semi-classical decomposition can be applied to other methods based on Feynman-diagrammatic approaches.
Therefore, the present work emphasizes the convenience and importance of the semi-classical expansion when we investigate QCD thermodynamics using non-lattice approaches.

In the lattice QCD simulation, the fluctuation of the Polyakov loop phase is explicitly and automatically included via the configuration generation process.
Thus, the lattice QCD simulation can reproduce RW transition without any problems.
This point is also discussed in the context of dual condensate (dressed Polyakov loop)~\cite{Bruckmann:2011zx,Bilgici:2009tx}.
In the calculation of the dual condensate, we must break RW periodicity and transition to make the dual condensate finite; for example, the lattice QCD data by fixing the Polyakov loop phase is one possibility.
In contrast, such treatment is not needed in simple-type Dyson-Schwinger equations because there is no RW periodicity.
This is the crucial difference induced by the existence and absence of RW periodicity and transition.

Since RW periodicity and the transition affect the thermodynamics not only at finite $\mu_\mathrm{I}$, but also at finite $\mu_\mathrm{R}$, we can discuss the additional potential term induced by the classical field in the future.
The canonical partition function is very sensitive to the property, and thus the Fourier coefficient may have important information about QCD.
Actually, there are some discussions on the QCD phase transition from the Fourier coefficient~\cite{Kashiwa:2017swa,Almasi:2019bvl,Vovchenko:2019hbc}.

\subsection{Polyakov-loop phase contributions}
At moderately high $T$, the Polyakov loop is an important quantity because it is not exactly one, thus the contributions of the phase of the Polyakov loop should be seriously considered.
In fact, the relation between the fluctuation of the Polyakov loop, which is somewhat related to its phase, and the deconfinement transition is known~\cite{Lo:2013hla}.
Furthermore, the distribution of the Polyakov loop phase in the color space may be important in the sense of partial deconfinement~\cite{Hanada:2023rlk}.
Since the absence of RW transition should be closely related to the Polyakov-loop phase contributions, it indicates that the absence of RW transition is related to the Polyakov-loop fluctuation.
The deviation from exact results may be related to the fluctuation of the Polyakov-loop phase: If the phase fluctuation is strongly suppressed, the deviation should be suppressed.
Intuitively, the suppression becomes stronger and stronger with increasing $T$ and thus the present discussions may be relevant at moderately high $T$, but it will join the game when $\mu$ becomes significantly large enough.
Therefore, at sufficiently high $T$ (with $\mu/T \ll 1$), any Feynman diagrammatic approach can reproduce exact results because of the strong suppression of the fluctuation of the Polyakov loop phase.
It should be noted that the Feynman diagrammatic approach without RW periodicity may have a large systematic error even at moderately high $T$ and then the deviation from the exact result induced by the absence of RW periodicity may be covered by the error margins.
Although this is true, it is important to know all possible origins of the deviation.

\section{Summary}
\label{Eq:Summary}

In this paper, we discuss the reliability of analytical methods for QCD at finite temperature ($T$) and real chemical potential ($\mu_\mathrm{R}$) based on the canonical approach.
In particular, we discuss the moderately high $T$ region where the Polyakov loop does not reach one; it corresponds to the region where the semi-quark-gluon plasma is realized.
Actually, we discuss the role of Roberge-Weiss (RW) periodicity, which sometimes loses in the non-lattice calculations, in thermodynamics.

In the canonical approach, we start from the grand canonical partition function at finite imaginary chemical potential ($\mu_\mathrm{I}$) and construct the canonical partition function with a fixed quark number. 
Since the process can be reversed, we can say that the analytical method, which cannot reproduce QCD properties at finite $\mu_\mathrm{I}$, such as RW periodicity and RW transition, has a deviation from the exact results.
The RW non-periodic approaches lead to a deviation at moderately high $T$ if it cannot reproduce RW periodicity and transition; it usually occurs.
The mismatch in the results is caused by extra contributions, that appear in the RW non-periodic approaches; it can be clearly seen from Eq.\,(\ref{eq:enq}) and is demonstrated in Appendix\,\ref{sec:examples}.
In particular, the deviation becomes serious when $\mu_\mathrm{R}/T$ increases because the fugacity makes the higher-order oscillation mode enhance at finite $\theta$; see Eq.\,(\ref{Eq:fugacity}).
This deviation should have a relationship with the fluctuation of the Polyakov-loop phase.

The absence of RW transition comes from the absence of $\mathbb{Z}_{N_\mathrm{c}}$ replicas where $N_\mathrm{c}$ is the number of colors.
It is natural to think that the absence of replicas is related to the incomplete treatment of the contributions of the Polyakov loop phase in the partition function.
Thus, this problem can be overcome by suitably introducing the Polyakov loop phase into the analysis.
One possible way is to decompose the gluon field into the sum of the classical field and the fluctuation, and after that loop diagrams are estimated with a suitable setup of the distribution of the classical field; it is called a semi-classical expansion.
Then, the Fourier coefficient can be used to check the validity of the additional potential term induced by the methods.

In this paper, we discuss contributions not only the absolute value of the Polyakov loop, but also the phase of the Polyakov loop.
Since the imaginary chemical potential has several important pieces of information about QCD, it indicates that we may be able to use the region to pick up several features of QCD based on machine learning and topological data analysis; this may be useful for feature extraction.
We hope that this discussion sheds some light on the properties of the semi-quark-gluon plasma.

\begin{acknowledgments}
This work is supported in part by Grants-in-Aid for Scientific Research from JSPS (No. JP22H05112).
\end{acknowledgments}

\appendix

\section{Polyakov loop computation}
\label{sec:A_PL}

One possible way to compute the Polyakov loop in the non-lattice approaches is that we introduce the $\mathbb{Z}_3$ symmetry breaking external field into Lagrangian and take the zero limit of the external field at the final stage.
In this section, we explain the way with the external field.

To introduce the external field, which is composed of the Polyakov loop, it must be imposed that the external field breaks the $\mathbb{Z}_3$ symmetry but does not break the extended $\mathbb{Z}_3$ symmetry due to the RW periodic nature of QCD; RW periodicity disappears if the extended $\mathbb{Z}_3$ symmetry is broken explicitly.
The simplest form of the external field term ${\cal L}_\mathrm{ext}$ in the Lagrangian density is
\begin{align}
    {\cal L}_\mathrm{ext}
    &= J^\mathrm{R}_\mathrm{ext} \Re \, \Phi 
      + J^\mathrm{I}_\mathrm{ext} \Im \, \Phi,
\end{align}
where we decompose the combination of $\Phi$ and ${\bar \Phi}$ to the real and imaginary parts.
Unfortunately, this form is not good for RW periodicity, and thus we should replace it as
\begin{align}
    {\cal L}_\mathrm{ext}
    &= J^\mathrm{R}_\mathrm{ext} \Re ( e^{i \theta} \Phi )
      + J^\mathrm{I}_\mathrm{ext} \Im ( e^{-i \theta} \Phi )
    \nonumber\\
    &= J^\mathrm{R}_\mathrm{ext} \Re \, \Psi 
      + J^\mathrm{I}_\mathrm{ext} \Im \, \Psi,
\label{eq:mPL}
\end{align}
where $\Psi=\exp(i\theta) \Phi$ is the so called modified Polyakov loop~\cite{Sakai:2008ga}; this form was first discussed in Ref.~\cite{Kashiwa:2009zz} in the different context.
Since the modified Polyakov loop and its conjugate are a RW periodic quantity, unlike the Polyakov loop, the form (\ref{eq:mPL}) is suitable from the point of view of RW periodicity.
After including the external term, we should finally take the $J^\mathrm{R,I}_\mathrm{ext} \to 0$ limit in the evaluation of observables;
\begin{align}
    \langle \Re \, \Psi \rangle &= \frac{1}{\beta V} \frac{\partial}{\partial J_\mathrm{ext}^\mathrm{R}} {\ln \cal Z}
    \Bigl|_{J_\mathrm{ext}^\mathrm{R} \to 0},
    \nonumber\\
    \langle \Im \, \Psi \rangle &= \frac{1}{\beta V} \frac{\partial}{\partial J_\mathrm{ext}^\mathrm{I}} {\ln \cal Z}
    \Bigl|_{J_\mathrm{ext}^\mathrm{I} \to 0}.
\end{align}
This is the standard procedure for investigating the phase transition.
It should be noted that we can evaluate $\Phi$ from $\Psi$~\cite{Kashiwa:2019dqn}.
With the procedure, we can individually control the $\mathbb{Z}_3$ symmetry and the extended $\mathbb{Z}_3$ symmetry, and then the numerical control of the limit $J_\mathrm{ext} \to 0$ may become easier.
Therefore, we should care for not only the classical field, but also the external field.

\section{RW periodic and non-periodic results}
\label{sec:examples}

In this Appendix, we consider the situation in which RW periodicity is introduced according to Eq.\,(\ref{eq:extension}) but the additional term is neglected in the sense of the semi-classical expansion.
This situation is demonstrated by using the Polyakov-loop extended Nambu--Jona-Lasinio (PNJL) model~\cite{Fukushima:2003fw}.
The PNJL model is constructed by combining the Nambu--Jona-Lasinio model and the Polyakov-loop potential, which is phenomenologically constructed.
The mean-field approximation for the PNJL model corresponds to the semi-classical expansion.
It should be noted that the Polyakov loop ($\Phi$) in the expansion should be larger than the correct Polyakov loop because of the Jensen inequality as
\begin{align}
    \Phi
    &= \frac{1}{N_\mathrm{c}} \Bigl \langle \mathrm{tr} \, {\cal P} \exp\Bigl( i g \int_0^\beta {\cal A}_0 \, d\tau \Bigr) \Bigr\rangle
    \nonumber\\
    &\le \frac{1}{N_\mathrm{c}}\mathrm{tr} \, e^{ig \langle {\cal A}_0^a \rangle},
         \label{eq:Polyakov_loop}
\end{align}
where $\tau$ is the imaginary time and ${\cal P}$ is the path ordering operator for the $\tau$ direction; for example, see Ref.\,\cite{Braun:2007bx}. 
It is noted that both sides of Eq.\,(\ref{eq:Polyakov_loop}) at purely real chemical potential in realistic QCD are real; for example, see Refs.\,\cite{Nishimura:2014rxa,Tanizaki:2015pua}.
If the non-trivial $\mathbb{Z}_3$ sector is realized at purely imaginary chemical potential, we should take the absolute value of both sides of Eq.\,(\ref{eq:Polyakov_loop}).
The first line in Eq.\,(\ref{eq:Polyakov_loop}) is the exact Polyakov loop, and the second shows the upper bound of the value.
In the mean-field approximation for the PNJL model, $\langle {\cal A}_0^a \rangle$ is considered the classical field contribution under suitable gauge fixing.
This means that we can prepare the situation that RW periodicity exists but the additional potential term is neglected by dropping the Polyakov-loop potential from the PNJL model; for details of the Polyakov-loop potential, see Refs.~\cite{Fukushima:2003fw,Ratti:2005jh,Roessner:2006xn} as an example.

\begin{figure}[t]
 \centering
 \includegraphics[width=0.48\textwidth]{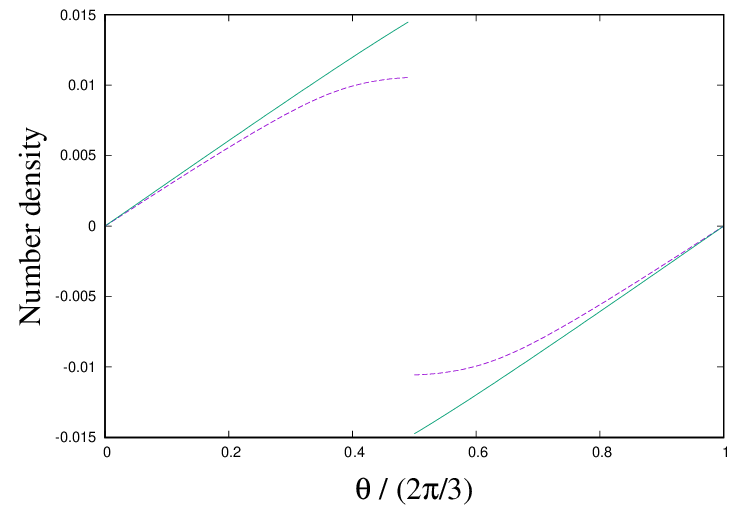}
 \caption{The quark number density as a function of $\theta$ at $T=300$ MeV.
          The dashed and solid lines show the result of the original PNJL model and the model without the Polyakov-loop potential.}
\label{fig:fig1}
\end{figure}

Figure\,\ref{fig:fig1} shows the $\theta$-dependence of the quark number density ($n_\mathrm{q}$) defined as
\begin{align}
    n _\mathrm{q} &= \frac{1}{\beta V} \frac{\partial}{\partial \mu} {\ln \cal Z}_\mathrm{PNJL},
\end{align}
where $V$ is the three-dimensional volume and ${\cal Z}_\mathrm{PNJL}$ is the grand canonical partition function of the PNJL model.
In the figure, the original PNJL model (dashed line) and the model without the Polyakov-loop potential (solid line) are shown, and then $T$ is set as $300$ MeV which is just an example, where the realization of RW transition is usually expected.
For the formulation and the parameter set used here, see Ref.\,\cite{Kashiwa:2019dqn} and reference therein.
The gap of the quark number density at $\theta = \pi/3$ indicates RW transition; the quark number density is purely imaginary at finite $\theta$.
In the figure, we can clearly see that the line in $0 \le \theta \le \pi/3 - \epsilon$ corresponds to the trivial $\mathbb{Z}_3$ sector, and the line in the $\pi/3 +\epsilon \le \theta \le \pi - \epsilon $ region corresponds to the result of another $\mathbb{Z}_3$ replica where $\epsilon$ is the infinitesimal value.
We can clearly see that there is a deviation between two models.
This difference becomes the origin of the difference of the Fourier coefficient.
This may mean that the inclusion of Eq.\,(\ref{eq:extension}) makes the deviation from the exact result milder because RW periodicity is restored, but there should still be a deviation.

In addition, at least in the PNJL model, RW transition still appears at low $T$ if the Polyakov loop potential is neglected; this problem is cured by including the potential term.
This means that not only in the high $T$ region, but also in the low $T$ region, the deviation from the exact results appears even if we extend the Matsubara frequency, and we must carefully consider the additional potential term.

\bibliography{ref.bib}

\end{document}